\documentclass[pra,twocolumn,amsmath]{revtex4}
\begin{document}

\title{Note on ``Hyperfine coherence in the presence of spontaneous photon scattering'' (quant-ph/0502063) and minimum energy requirements for quantum logic}
\author{Julio Gea-Banacloche}
\email[]{jgeabana@uark.edu}
\affiliation{Department of Physics, University of Arkansas, Fayetteville, AR 72701}

\date{\today}                                          

\begin{abstract}
The goal of this short note is to show that the formulas I derived originally in \cite{me1} regarding the errors introduced in quantum logical operations by the quantum nature of the control fields apply even in the situation discussed recently by Ozeri et al. in \cite{ozeri}, where the decoherence-inducing spontaneous Raman scattering is considerably suppressed. 
\end{abstract}

\maketitle

\section{Introduction}

A few years ago I argued \cite{me1} that the quantum fluctuations in a laser field used to do quantum logic on an atomic system would introduce an error probability per operation that scaled, for a coherent state, as the inverse of the total number of photons in the pulse.  This result was in agreement with results obtained independently, at around the same time, by other authors, using a wide range of diverse methods \cite{barnes,vanenk,ozawa}.  

Not long afterwards, Itano \cite{itano} argued, on the basis of the Mollow transformation \cite{mollow}, that, for a coherent-state pulse, the only decoherence should be that arising from spontaneous emission.  A spirited discussion followed \cite{vanenk2,me2}, but eventually the dust settled and the various results were shown to be quite compatible:  in particular, for an ordinary, dipole-allowed atomic transition in free space, $1/n$ was shown \cite{me2} to be of the order of the spontaneous emission probability into the solid angle of the incident beam, which is, of course, always less than the total spontaneous emission rate, and hence a proper lower bound to the total error probability claimed by Itano.  In the process, a tighter lower bound was found, that could be written as $1/n'$, where $n'$ is not the total number of photons in the pulse but only the number within an area $\sigma$ equal to the total resonant scattering cross-section for an atomic dipole transition.  The ratio $1/n'$ turned out to be equal, up to numerical factors of the order of unity, to the total probability of spontaneous emission in \emph{any} direction during the time of a nontrivial logical operation (e.g., a bit flip, or $\pi$ pulse).  Thus, if $\epsilon$ is the total error probability for such an operation, one generally would have $\epsilon\ge 1/n' \ge 1/n$.

Meanwhile, in an unpublished note \cite{me3}, I had argued that the same constraint would apply to a quantum logical operation that proceeded via a Raman, two-photon process. In a very interesting, recent paper \cite{ozeri}, however, the Boulder group has shown that when the Raman transition involves two upper levels both about equally detuned from resonance (that is, the spacing between upper levels is $\Delta_f$ and the single-photon detuning is $\Delta \gg \Delta_f$), the state-changing, or Raman, part of the spontaneous emission is considerably suppressed, and the total decoherence of the ion is proportional to only this fraction of the spontaneous emission, so it is considerably suppressed as well.  

Under these circumstances, it is natural to ask if the original estimates \cite{me1,barnes,vanenk,ozawa} for the error probability as a function of the total number of photons in the pulse still apply.  I show below that they do.  Caveat: none of what follows attempts to track down very carefully numerical factors of the order of 2. 

\section{Error probability as a function of $n$ for the setup in \cite{ozeri}}

According to Ozeri et al., \cite{ozeri}, in their setup the effective Rabi frequency for the two-mode transition is reduced by the same interference mechanism that reduces the Raman scattering rate, so that it now scales as $1/\Delta^2$ for large $\Delta$.  I take this to mean that the ``ordinary'' effective Rabi frequency for a two-photon Raman transition, proportional to $g^2/\Delta$, is reduced by a factor $\Delta_f/\Delta$, so
\begin{equation}
\Omega_\text{eff} = \frac{g^2}{\Delta}\,\frac{\Delta_f}{\Delta} =  \frac{g^2 \Delta_f}{\Delta^2}
\label{e10}
\end{equation}
(I attempt to use the notation of \cite{ozeri} throughout; see below for an explicit expression for $g$).
 
Ozeri et al. also state that the total number of photons scattered during a $2\pi$ pulse goes as $2\pi\gamma/\Delta_f$, whereas the spontaneous Raman scattering over the same time goes down as $1/\Delta^2$.  I take this to mean that the probability to emit a spontaneous Raman (state-changing) photon during a $\pi$ pulse is proportional to 
\begin{equation}
\epsilon = \frac{\pi\gamma}{\Delta_f}\left(\frac{\Delta_f}{\Delta}\right)^2
\label{e11}
\end{equation}
and I identify this (up to a factor of the order of unity) with the gate error probability for this bit-flip operation.  Since the time, $T$, for the $\pi$ pulse must satisfy $\Omega_\text{eff}T = \pi/2$, we clearly have, from Eq.~(\ref{e10}),
\begin{equation}
\frac{\Delta_f}{\Delta^2} = \frac{\pi}{2 g^2 T}
\label{e12}
\end{equation}
and using this in Eq.~(\ref{e11})
\begin{equation}
\epsilon = \frac{\pi^2 \gamma}{2 g^2 T}
\label{e13}
\end{equation}
or introducing $g = Ed/2\hbar$ (where $E$ is the laser field amplitude and $d$ the $S\to P$ transition's atomic dipole moment), and $\gamma = (1/4\pi\epsilon_0)(4\omega^3 d^2/3\hbar c^3)$, as typical Rabi frequencies and radiative decay rates for an electric-dipole allowed transition, we get
\begin{equation}
\epsilon = \frac{8 \pi^3}{\lambda^2}\,\frac{\hbar\omega}{3 \epsilon_0 c E^2 T}
\label{e14}
\end{equation}Finally, note that according to \cite{ozeri}, $E^2 = 2 I/c\epsilon_0$, where $I$ is the laser intensity (power per unit area), and introduce $\sigma = 3\lambda^2/2\pi$, the total resonant scattering cross-section for an atomic dipole transition.  Eq.~(\ref{e14}) can then be written as
\begin{equation}
\epsilon = 2\pi^2\,\frac{\hbar\omega}{I\sigma T} \ge \frac{1}{n'}
\label{e15}
\end{equation}
where $n'$ is the number of laser photons in the volume $\sigma c T$, and can be interpreted as the number of laser photons that interact effectively with the atom.  Since $\sigma<\lambda^2$, for any laser pulse of duration $T$ focused to a spot larger than one wavelength in radius, the total number of photons in the pulse, $n$, will be larger than $n'$, so we still have
\begin{equation}
\epsilon \ge \frac{1}{n'} \ge \frac{1}{n}
\label{e16}
\end{equation}
as a bound on the achievable error in a quantum logical gate that uses a control field with $n$ photons, as in my original paper \cite{me1}, and in the Reply \cite{me2}.

\section{Discussion}

The suppression of decoherence illustrated in \cite{ozeri} is certainly very interesting, and the authors' analysis is very illuminating, but the calculations in the previous section show that, when it comes to doing quantum logic, one still essentially has to ``buy'' the increased fidelity at the expense of increased energy in the control field, in agreement with the prediction already made in \cite{me1}, and the tighter limit derived in \cite{me2}.   It is, therefore, justified to consider this as yet another instance of the validity of the result, concerning energy requirements for quantum logic, which I first postulated generally in \cite{me4}, and for which, to date, no counterexamples have been found.  

A simple way to understand the idea behind \cite{me4} is to consider that, at a minimum, the qubit system has to couple to the control system, and if the latter is quantum in nature this means the qubit couples to the control's quantum fluctuations, as well as to its deterministic part.  Under these circumstances, the maximum achievable fidelity must be limited by the signal-to-(quantum) noise ratio of the control; for a control in a coherent state, in appropriate (energy) units, this ratio is $n$:1, leading to the result (\ref{e16}).  

I would like to stress that this concept of ``decoherence induced by the quantum nature of the control system'' is a perfectly legitimate one; it just, by its very nature, tends to underestimate the total decoherence.  The recent work of Silberfarb and Deutsch \cite{silber} can be used to make the point.  They find that the effective single-mode (Jaynes-Cummings-type) approaches to the atom-laser field interaction \cite{vanenk} actually predict the correct amount of atom-field entanglement, even for an atom in free space interacting with a paraxial beam, as long as the total decoherence rate for the system remains small.  This entanglement (which clearly arises from the quantum nature of the field!) would, by tracing over the field state, result in a decoherence of the atomic qubit of the order of $1/n$.  The subtle point (already conceded in \cite{me2}, and recently independently confirmed by the analysis of \cite{nha}) is that, for a field in a coherent state, this is not {\em in addition to}, but rather implicitly contained in, the total decoherence rate that one would calculate, using the Mollow transformation, for the same system, and that one would, in this picture, attribute to spontaneous emission only.

We have, therefore, two different, overlapping approaches, each with its strengths and weaknesses.  The Mollow transformation, of course, does not apply if the field is not in a coherent state, but where it applies it leads more naturally to the tighter limit $\epsilon > 1/n'$ (which does not suffer from the ``area paradox'' originally pointed out by Itano \cite{itano}).  The ``control-induced decoherence'' viewpoint, on the other hand, makes it easier to see that the limit $\epsilon > 1/n$ must still apply, even in situations like the one considered in this note, for which, in the words of \cite{ozeri}, ``the total scattering rate gives a pessimistic measure of decoherence.''    
   
Finally, I should point out that M. Ozawa and I have recently shown \cite{JGBandOzawa} that his ``conservation-law induced quantum limit,'' or CQL (see \cite{ozawa,ozawa2}), holds, in the rotating-wave approximation, for a wide range of atomic systems, including multilevel, Raman-coupled atoms interacting with a multimode vacuum.  From this we can show quite rigorously that, for coherent-state fields, a scaling of the infidelity with $n$ of the form $\sim 1/n$ must hold in all these systems.  We also show in \cite{JGBandOzawa} (keeping very careful track, this time, of all the factors of 2) that this CQL is related, in a certain limit, to the field's \emph{phase} fluctuations only.  While this means that the CQL is not, typically, the tightest possible limit of the form (\ref{e16}) that may apply in a particular situation, it is, however, as far as we can tell, an inescapable one.

\end{document}